\documentclass[pss,fleqn,embeddedheads]{w-art}
\usepackage{times}
\usepackage{w-thm}
\usepackage[]{graphicx}
\begin{document}
\DOIsuffix{theDOIsuffix}
\Volume{XX}
\Issue{1}
\Copyrightissue{01}
\Month{01}
\Year{2004}
\pagespan{1}{}
\Receiveddate{\sf zzz} \Reviseddate{\sf zzz} \Accepteddate{\sf
zzz} \Dateposted{\sf zzz}



\title[Strongly correlated multilayered nanostructures near the Mott transition]
{Strongly correlated multilayered nanostructures near the Mott transition}


\author[J. K. Freericks]{J. K. Freericks\footnote{Corresponding
     author: e-mail: {\sf freericks@physics.georgetown.edu}, Phone: +\,202\,687\,6179, Fax:
     +\,202\,687\,2087}}
\address[]{Department of Physics, Georgetown University, Washington, DC
20057, U.S.A.}
\begin{abstract}
We examine devices constructed out of multilayered sandwiches of
semi-infinite metal--barrier--semi-infinite metal, with the barrier tuned to 
lie near the quantum
critical point of the Mott metal-insulator transition.  By employing
dynamical mean field theory, we are able to solve the many-body problem
exactly (within the local approximation) and determine the density
of states through the nanostructure and the charge transport perpendicular
to the planes.  We introduce a generalization of the Thouless energy
that describes the crossover from tunneling to incoherent
thermally activated transport.
\end{abstract}
\maketitle                   





\section{Introduction}

Many new technological developments are anticipated over the coming
years in the field of nanotechnology.  There is a current interest in trying
to incorporate strongly correlated materials into nanoscale devices, because
strongly correlated systems often have interesting bulk properties that
can be tuned by changing the pressure, temperature, chemical doping, etc.
What is lesser known is how the strong electron correlations are modified
when the bulk materials are confined on the quantum scale and attached
to other (noncorrelated) materials, like normal metallic leads. In particular,
we expect there to be a reorganization of the electronic states driven by
a charge transfer, associated with the mismatch of the chemical potentials,
and by the normal-state proximity effect of the metallic leads on the insulator,
which will produce exponentially decaying states within the barrier of the
Mott insulator, analogous to the superconducting proximity effect in normal
metals.  Here we adjust our system to be overall charge neutral, so we do
not investigate the charge transfer; we only investigate the proximity
effect.

From a device standpoint, the simplest type of nanostructure to create is
a multilayered structure, where stacks of planes of one material are topped
by another material, and so on until a given device and heterostructure
is made.  Recent advances in pulsed laser deposition and molecular
beam epitaxy have allowed many complex structures to be grown, with 
interfaces between the different materials being well defined up
to a few atomic layers.

The question we will address in this contribution is how do the properties
of semi-infinite metal--barrier--semi-infinite metal multilayered 
heterostructure vary with the
barrier thickness when the barrier is a Mott insulator, tuned to lie
just slightly above the Mott metal-insulator transition (i.e. very close
to the quantum-critical point, but on the insulating side).  We examine
both single-particle properties like the density of states (DOS), including
a proof that the local DOS in the barrier of a single-plane barrier
has metallic behavior at strong coupling.  We also investigate
transport, and discuss a generalization of the Thouless energy that is
appropriate for Mott-insulating systems and governs the crossover from tunneling
to incoherent transport. We employ dynamical mean field
theory (DMFT) to perform these calculations, which relies on the local
approximation for the self energy to be accurate in these three-dimensional
inhomogeneous systems.

\begin{vchfigure}[htbf]
\includegraphics[width=.650\textwidth]{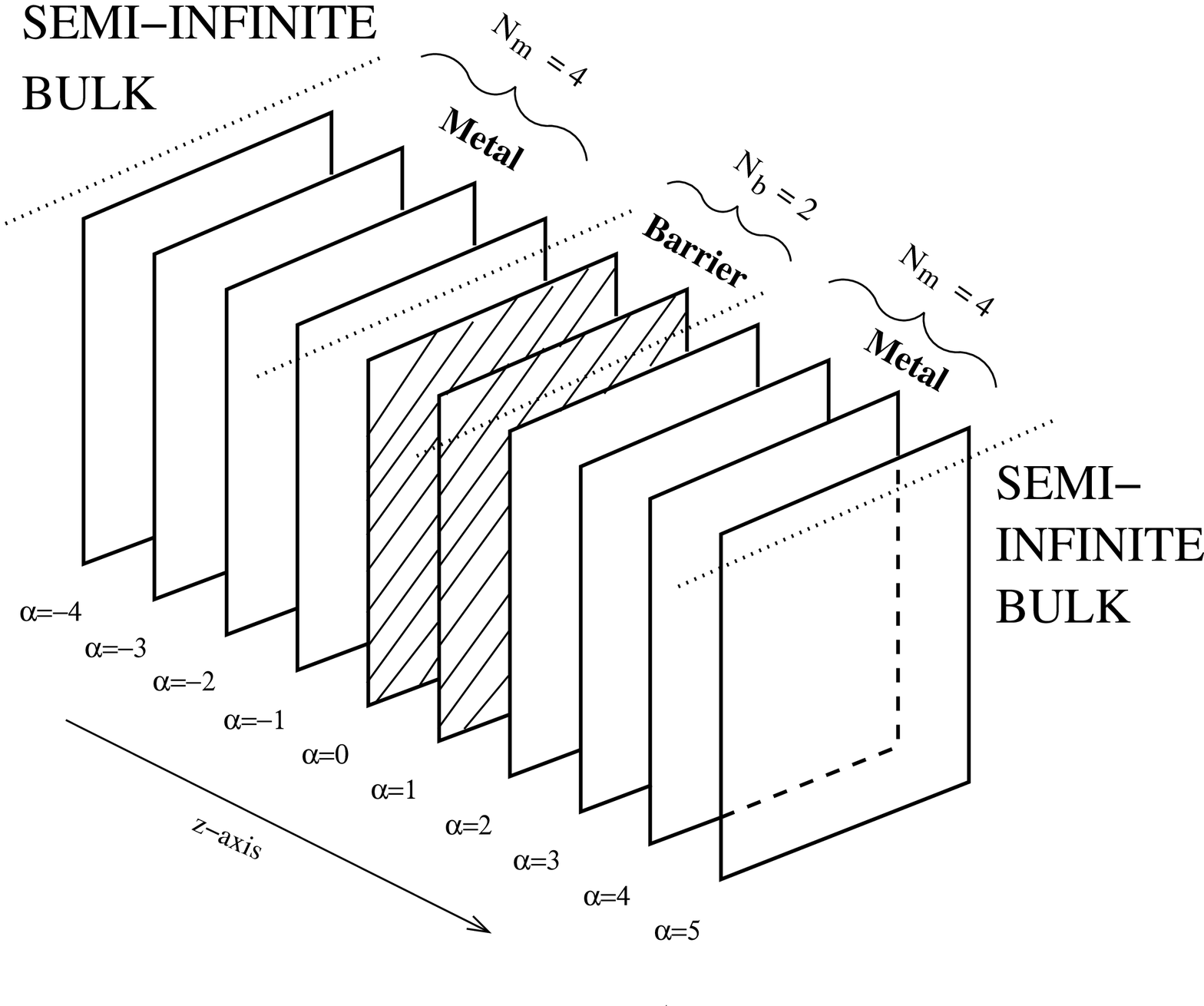}
\caption{Schematic of the multilayered nanostructure, where we take a finite
number of self-consistent metal planes ($N_m=4$ in the figure, but $N_m=30$
in our calculations), couple them on one end to a bulk semi-infinite metal,
and on the other end to a barrier described by the Falicov-Kimball model
(with $N_b=2$ in the figure depicted by the cross-hatched planes; in our
calculations $N_b=N$ varies from 1 to 30).
} \label{fig:0}
\end{vchfigure}

\section{Formalism}

We describe the Mott insulator by the spinless Falicov-Kimball 
model~\cite{falicov_kimball_1969}
\begin{equation}
\mathcal{H}=-\sum_{ij}t_{ij}c^\dagger_ic_j+\sum_iU_i\left (
c^\dagger_ic_i- \frac{1}{2}\right )\left ( w_i-\frac{1}{2}\right )
\label{eq: hamiltonian}
\end{equation}
where $t_{ij}$ is a Hermitian hopping matrix, $U_i$ is the Falicov-Kimball
interaction, and $w_i$ is a classical variable that equals one if
there is a localized particle at site $i$ and zero if there is no
localized particle at site $i$ (a chemical potential $\mu$ is employed
to adjust the conduction-electron concentration).  Since we are considering
multilayered
heterostructures (see Fig.~\ref{fig:0}), we assume that 
the hopping matrix is translationally invariant
within each plane, as well as the Falicov-Kimball interaction. For simplicity,
we will take the lattice sites to lie on the sites of a simple cubic lattice,
with $t_{ij}=t$ for all nearest neighbors (we use $t$ as our unit of energy), 
and we will take $U_i=U$ for all
lattice sites $i$ that lie within the barrier plane.  This choice assumes that
the bare kinetic energy is the same for the metallic leads and for the 
barrier, and that the barrier is strongly correlated via the Falicov-Kimball
interaction (renormalizing its bandstructure) within the barrier planes.  We 
also work at half filling,
with $\mu=0$ and $\langle w_i\rangle = w_1=1/2$.  In this case, the chemical
potential has no temperature dependence, and the electronic charge
remains homogeneous throughout the system.

The dynamical mean field theory for inhomogeneous systems was originally
worked out by Potthoff and Nolting~\cite{potthoff_nolting_1999} and
developed for these particular heterostructures in another 
publication~\cite{freericks_unpub}.  Here we just include the relevant 
summarizing formulas.
The starting point is to note that the system has translational invariance
in the two-dimensional planar direction, so we can Fourier transform
from real space to momentum space; all physical quantities we will
be interested in here depend only on the two-dimensional bandstructure
$\epsilon^{2d}=-2t[\cos{\bf k}_x+\cos{\bf k}_y]$.  We let a Greek letter
($\alpha$, $\beta$, $\gamma$, ...) denote the $z$-component of each of
the stacked planes. Then, because an electron with energy $\epsilon^{2d}$
in the perpendicular direction, decouples from electrons with different
perpendicular energy, the problem for the Green's function reduces to
a quasi-one-dimensional problem, that can be solved with the
renormalized perturbation expansion~\cite{economou_1983}.  The result
for the local retarded Green's function at plane $\alpha$ is
\begin{equation}
G_{\alpha}(\omega)=\int d\epsilon^{2d}\rho^{2d}(\epsilon)
\frac{1}{L_{\alpha}(\epsilon^{2d},\omega)+
R_{\alpha}(\epsilon^{2d},\omega)-[\omega+\mu-\Sigma_\alpha(\omega)
-\epsilon^{2d}]}
\label{eq: g_loc}
\end{equation}
with $\rho^{2d}$ the DOS of a two-dimensional tight-binding square lattice
and $\Sigma_\alpha(\omega)$ the self energy at plane $\alpha$.  The
functions $R$ and $L$ are determined via recursion relations:
\begin{eqnarray}
R_{\alpha+n}(\epsilon^{2d},\omega)&=&\omega+\mu-\Sigma_{\alpha+n}(\omega)-
\epsilon^{2d}-
\frac{1}
{R_{\alpha+n+1}(\epsilon^{2d},\omega)}; 
\label{eq: r_recurrence}\\
L_{\alpha-n}(\epsilon^{2d},\omega)&=&\omega+\mu-\Sigma_{\alpha-n}(\omega)-
\epsilon^{2d}-
\frac{1}
{L_{\alpha-n-1}(\epsilon^{2d},\omega)};
\label{eq: l_recurrence}
\end{eqnarray}
for $n>0$. These recurrences are solved by starting at $n=\pm\infty$ for the
right or left recurrence, and then iterating in $n$.  Of course, in real
calculations, we must assume $R_\alpha=R_\infty$ and $L_\alpha=L_{-\infty}$
for all $\alpha$ up to a finite distance away from the interfaces with
the barriers; we include 30 such self-consistent planes in the metallic
leads (on each side of the barrier) in our calculations.
We determine $R_{\infty}$ ($L_{-\infty}$) by substituting $R_{\infty}$
($L_{-\infty}$)
into both the left and right hand sides of Eq.~(\ref{eq: r_recurrence})
[Eq.~(\ref{eq: l_recurrence})],
which produces a quadratic equation that  is solved by
\begin{equation}
R_{\infty}(\epsilon^{2d},\omega)=
\frac{\omega+\mu-\Sigma_{bulk}(\omega)-\epsilon^{2d}}{2}
\pm\frac{1}{2}\sqrt{[\omega+\mu-\Sigma_{bulk}(\omega)-
\epsilon^{2d}]^2-4}=L_{-\infty}(\epsilon^{2d},\omega),
\label{eq: recurrence_oo}
\end{equation}
with the sign of the square root chosen by analyticity or continuity.
In both Eqs.~(\ref{eq: r_recurrence}) and (\ref{eq: l_recurrence}),
we see that whenever the imaginary part of $R$
or $L$ is positive, it remains positive in the recursion, implying stability;
similarly, when the imaginary part is zero, we find the large root is
stable, which is the physical root.  Hence the recurrences are stable.

Once we have determined the local Green's function on each plane, we can 
perform the DMFT calculation to determine the local self
energy on each plane~\cite{brandt_mielsch_1989,freericks_zlatic_2003b}.
We start with Dyson's equation, which defines the effective medium for each 
plane
\begin{equation}
G_{0\alpha}^{-1}(\omega)=G_\alpha^{-1}(\omega)+\Sigma_\alpha(\omega).
\label{eq: dyson}
\end{equation}
The local Green's function for the $\alpha$th plane satisfies
\begin{equation}
G_\alpha(\omega)=(1-w_1)\frac{1}{G_{0\alpha}^{-1}(\omega)+\frac{1}{2}U}
+w_1\frac{1}{G_{0\alpha}^{-1}(\omega)-\frac{1}{2}U},
\label{eq: impurity}
\end{equation}
with $w_1$ equal to the average filling of the localized particles
[note that this above form is slightly different from the usual
notation~\cite{freericks_zlatic_2003b},
because we have made the theory particle-hole symmetric by the choice of
the interaction in Eq.~(\ref{eq: hamiltonian}), so that $\mu=0$
corresponds to half filling in the barrier region and in the ballistic
metal leads]. Finally, the self energy is found from
\begin{equation}
\Sigma_\alpha(\omega)=G_{0\alpha}^{-1}(\omega)-G_\alpha^{-1}(\omega).
\label{eq: sigma}
\end{equation}
The full DMFT algorithm 
begins by (i) making a choice for the self energy on each plane.  Next,
we (ii) use the left and right recurrences in Eqs.~(\ref{eq: r_recurrence}) and
(\ref{eq: l_recurrence}) along with the bulk values found in
Eqs.~(\ref{eq: recurrence_oo}) and the 30
self-consistently determined planes within the metal leads
to calculate the local Green's function at each plane
in the self-consistent region
from Eq. (\ref{eq: g_loc}).  Once the local Green's
function is known for each plane, we then (iii) extract the effective medium
for each plane from Eq.~(\ref{eq: dyson}), (iv) determine the new local
Green's function from Eq.~(\ref{eq: impurity}), and (v) calculate the
new self energy on each plane from Eq.~(\ref{eq: sigma}). Then we iterate
through steps (ii)--(v) until the calculations have converged.

\section{Numerical results}

The first thing we consider is the local DOS on the barrier plane for
a single-plane barrier ($N=1$) in Fig.~\ref{fig:1}~(a).  Note how the 
DOS looks like a Mott
insulator for large $U$ (in the bulk, the Mott transition occurs 
at $U\approx 4.9$), with an upper and lower Hubbard band forming,
but there is substantial low-energy DOS coming from the interface
localized states that are actually metallic in character (negative second 
derivative of the DOS with respect to $\omega$).  In addition, we show the
Friedel-like oscillations that develop in the metallic leads as we
move away from the barrier interface (for $U=5$) in Fig.~\ref{fig:1}~(b).
Note how the oscillations have a shorter period, and a smaller amplitude as
we move away from the interface. Already at 10 lattice spacings away from the 
interface, we see there is only a small difference from the bulk DOS
(which is why 30 self-consistent planes is sufficient for this work).

\begin{vchfigure}[htbf]
\centerline{
\includegraphics[width=.45\textwidth]{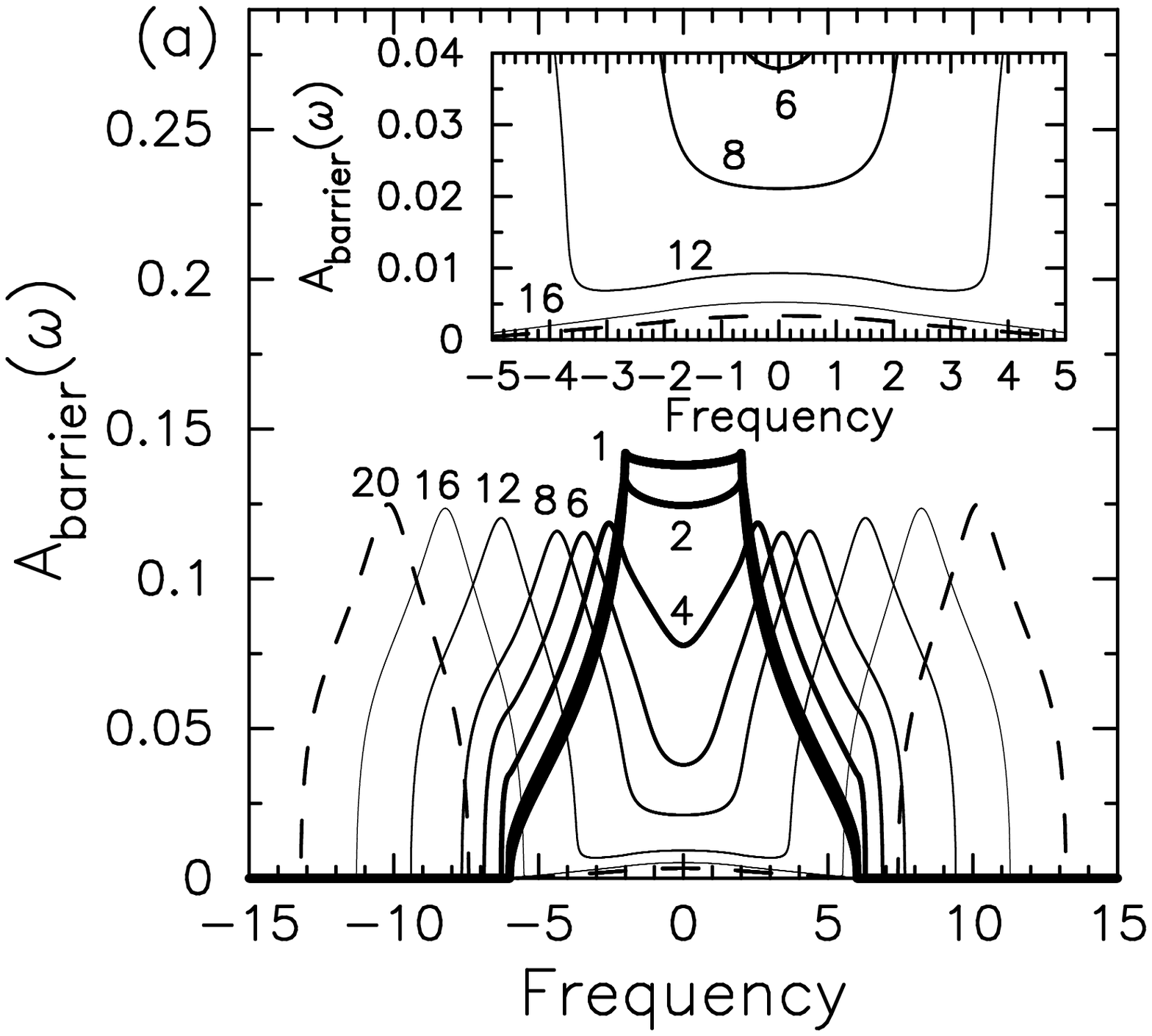}
\includegraphics[width=.45\textwidth]{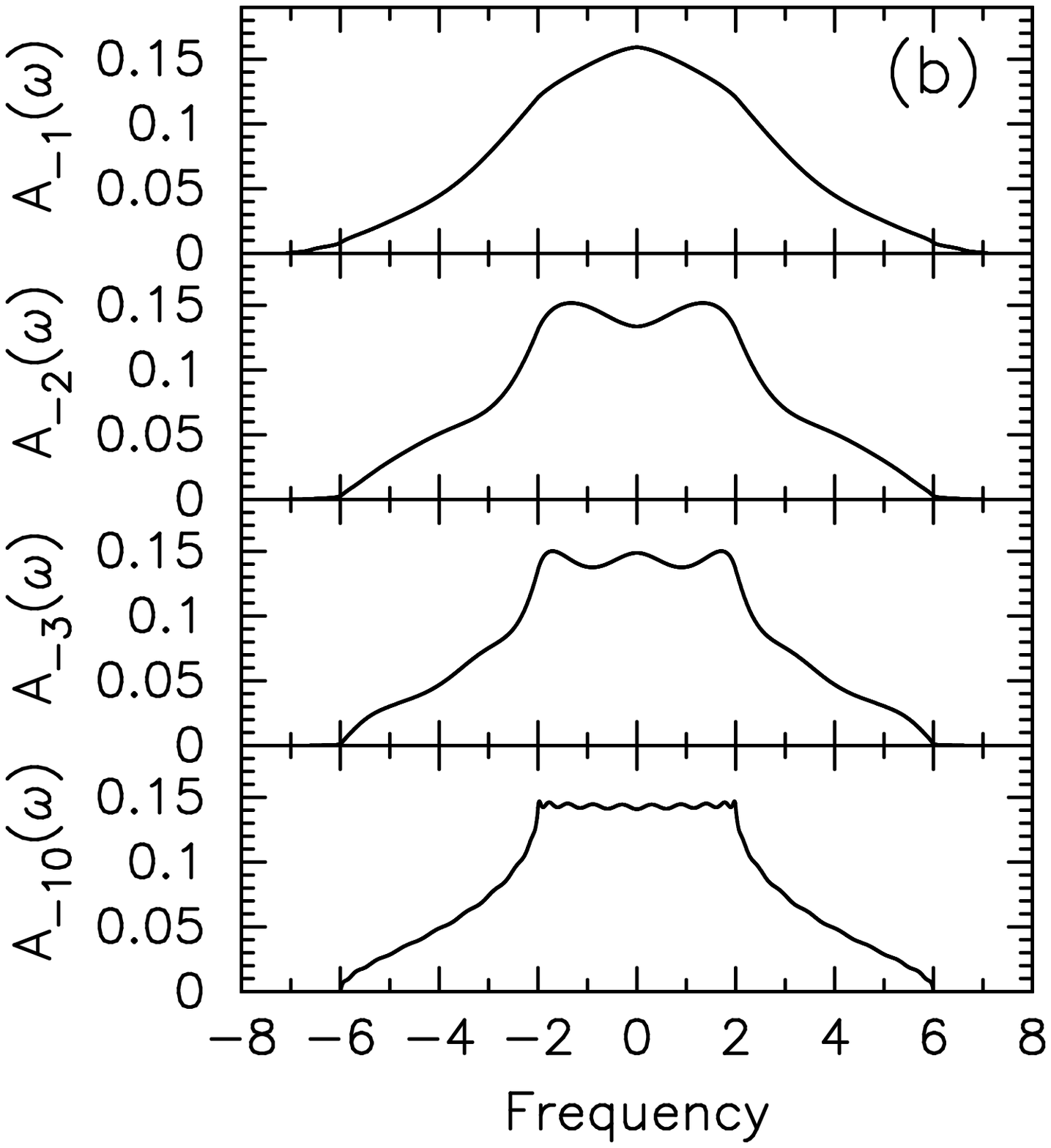}
}
\caption{(a) Local DOS for the barrier plane when $N=1$ and for various $U$
(indicated by the numerical labels).
The inset highlights the low-energy region, where one can see a low-weight 
metallic DOS form for large $U$ from the interface localized states of the
nanostructure. (b) The local DOS in  four barrier planes for $U=5$ and
$N=5$ (the first barrier plane is at $\alpha=0$; the metallic lead runs from
$\alpha=-1$ to $\alpha=-30$ on the left hand side).  Note how the amplitude 
of the Friedel-like oscillations are
quite small even by the time we hit the tenth plane from the barrier.
} \label{fig:1}
\end{vchfigure}

We show that the peak value at $\omega=0$ is easy to derive from an analysis
of the Potthoff-Nolting algorithm, and that it is nonvanishing for all
$U$.  To begin, we consider the non-self-consistent solution, where
we set $L_\alpha=L_{-\infty}$ and $R_{\alpha}=R_{\infty}$ for all
$\alpha$ except $\alpha=0$, where we have a Falicov-Kimball interaction.
Then the local Green's function in the barrier is
\begin{equation}
G_{\alpha=0}(\omega)=\int d\epsilon \rho^{2d}(\epsilon)\frac{1}
{2(\omega+\frac{1}{2}U-\epsilon)-\Sigma_0(\omega)\pm\sqrt{(\omega-\epsilon)^2-4}},
\label{eq: G_0}
\end{equation}
where the sign of the square root is chosen by either analyticity, or 
continuity. If we assume $\Sigma_{0}$ is large (which occurs in the 
Mott insulator) then we can expand Eq.~(\ref{eq: G_0}) in a power
series in inverse powers of $\Sigma$ to give
\begin{equation}
G_{\alpha=0}(\omega)\approx -\frac{1}{\Sigma_{0}(\omega)}-\frac{2\omega+U\pm 
s(\omega)} {\Sigma^2_{0}(\omega)}+...
\label{eq: g_expand}
\end{equation}
with $s(\omega)=\int \rho^{2d}(\epsilon) \sqrt{(\omega+\frac{1}{2}U-\epsilon)^2
-4}$. Taking
the value for $G_{\alpha=0}$ from Eq.~(\ref{eq: g_expand}), and plugging it into
the self-consistent DMFT algorithm described above, allows us to solve for
the self energy directly, with the result
\begin{equation}
\Sigma_{0}(\omega)=\frac{\frac{1}{4}U^2}{2\omega+U\pm s(\omega)}
\label{eq: sigma_large}
\end{equation}
which is large for $U\gg 1$, consistent with our ansatz. If we perform the
integral in the definition of $s(\omega)$, we find it satisfies
$s(\omega)=(\omega+\frac{1}{2}U)\cdot 0.653+i\cdot 1.05$.  Substituting this
result into the self energy, and then into the Green's function, and
evaluating the DOS, finally yields
\begin{equation}
\rho_0(\omega)\approx \frac{1.05}{\pi[(1.05^2+\frac{1}{4}U^2]};
\label{eq: dos_0}
\end{equation}
we compare this result to the exact calculated result in Table~\ref{table:1}.
One can see the agreement is excellent.

\begin{vchtable}[htb]
\vchcaption{Comparison of the results from Eq.~(\ref{eq: dos_0}) and the exact
numerical DOS for a single-plane barrier and various $U$.
} \label{tab:1}\renewcommand{\arraystretch}{1.5}
\begin{tabular}{lll} \hline
$U$&approximate result [Eq.~(\ref{eq: dos_0})]&exact numerical result \\
\hline
6&0.033&0.037\\
8&0.0195&0.021\\
12&0.009&0.009\\
16&0.005&0.005\\
20&0.003&0.003\\
\hline
\end{tabular}
\label{table:1}
\end{vchtable}

In Fig.~\ref{fig:2}, we plot a false color (grayscale) plot of the
DOS of the nanostructure in the near critical region $U=5$ for
a moderately thick ($N=20$) barrier.  Note how the Friedel oscillations
are most apparent in the center of the band in the metallic
leads (upper part of the graph).  We only plot the 40 planes on the 
left-hand-side of the nanostructure, because the symmetry of the structure
guarantees the rest of the nanostructure can be determined by a mirror
plane reflection.  Note how there are few oscillations within the
barrier itself, and how the DOS rapidly becomes small at low energy
(because it is an insulator). It is possible to even see some oscillations
induced in the metallic lead at energies close to the band edge, and 
at positions close to the interface.

\begin{vchfigure}[htbf]
\includegraphics[width=.75\textwidth]{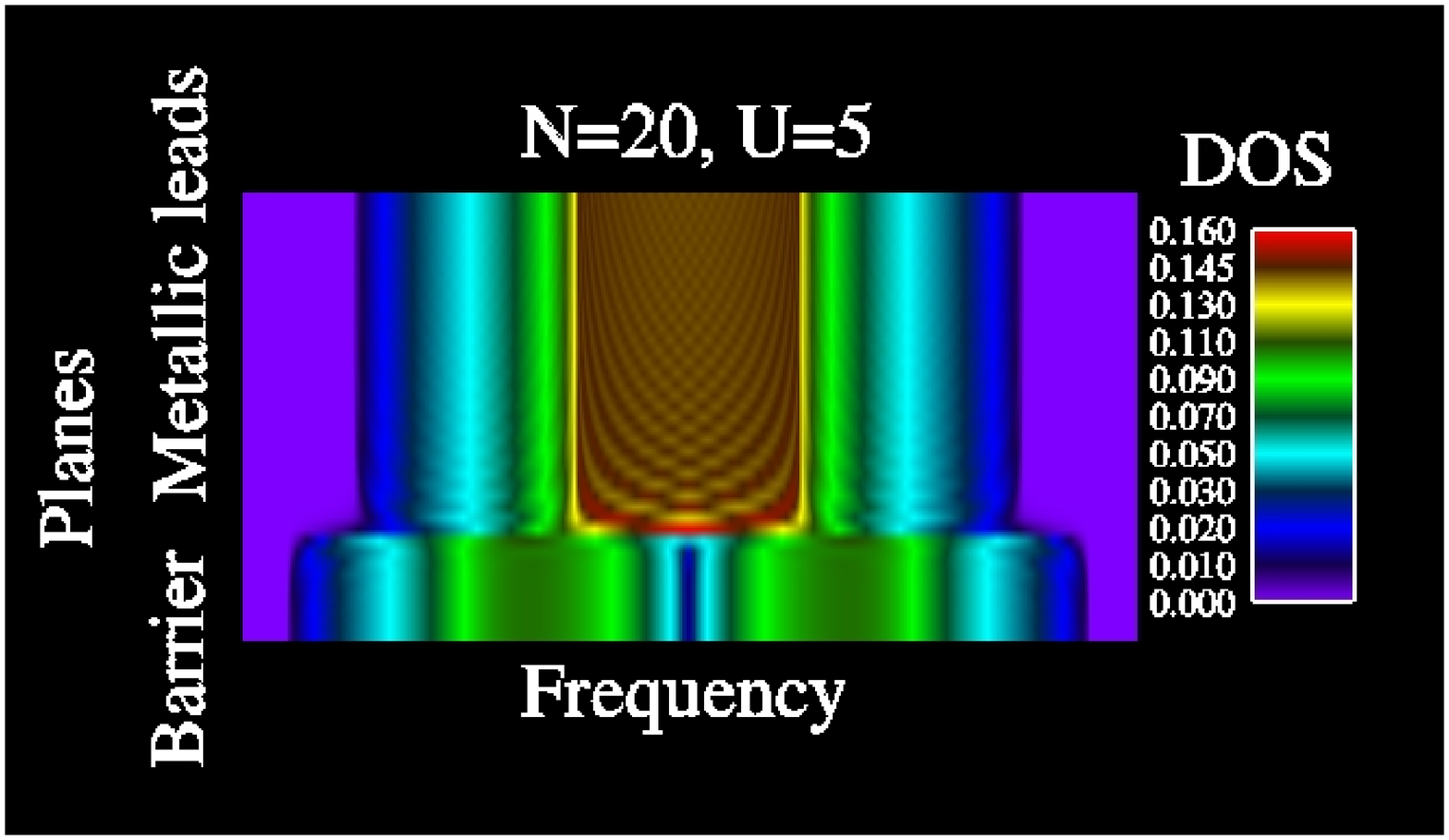}
\caption{False color (grayscale) plot of the local DOS for the near critical
$U=5$  nanostructure with $N=20$ planes in the barrier. Note how the
Friedel oscillations in the metallic lead are most apparent near the
center of the band, and how there are limited oscillations in the barrier
(the DOS decays exponentially fast with position in the barrier at low 
energy). (Color on-line.)
} \label{fig:2}
\end{vchfigure}

We examine transport properties in Fig.~\ref{fig:3}.  The resistance of a 
junction is calculated in the linear-response regime via a Kubo-based
formalism~\cite{kubo_1957} with the current-current correlation 
function~\cite{miller_freericks_2001,freericks_unpub}. The formalism requires
us to employ a conductivity matrix in real space, with matrix components
corresponding to the $z$-axis label of the different planes in the system.
The resistance-area product can be calculated for any temperature.

We extract an energy scale from the resistance, which we call the Thouless
energy~\cite{edwards_thouless_1972,thouless_1974}, since it reduces to the 
well-known diffusive and ballistic limits, but it also defines an energy 
scale when the barrier is a Mott insulator. In the insulating phase, it
is a function of time, and the point where $E_{Th}\approx T$
defines an important energy scale for the dynamics of the transport in
a nanostructure---it signifies when the transport crosses over
from tunneling behavior at low $T$ to incoherent ``Ohmic'' transport
at high $T$ [see Fig.~\ref{fig:3}~(a)]. On the other hand, the Thouless
energy is inversely proportional to $L$, the thickness of the barrier
for ballistic transport and inversely proportional to $L^2$ for
diffusive transport [this can be seen from taking the low-temperature
limit of Eq.~(\ref{eq: thouless}) and noting that the DOS is nearly constant
near the Fermi energy at low temperature and recalling the $R_n\propto L$ for
diffusive transport and $R_n$ is independent of $L$ for ballistic transport].

\begin{vchfigure}[htbf]
\centerline{
\includegraphics[width=.435\textwidth]{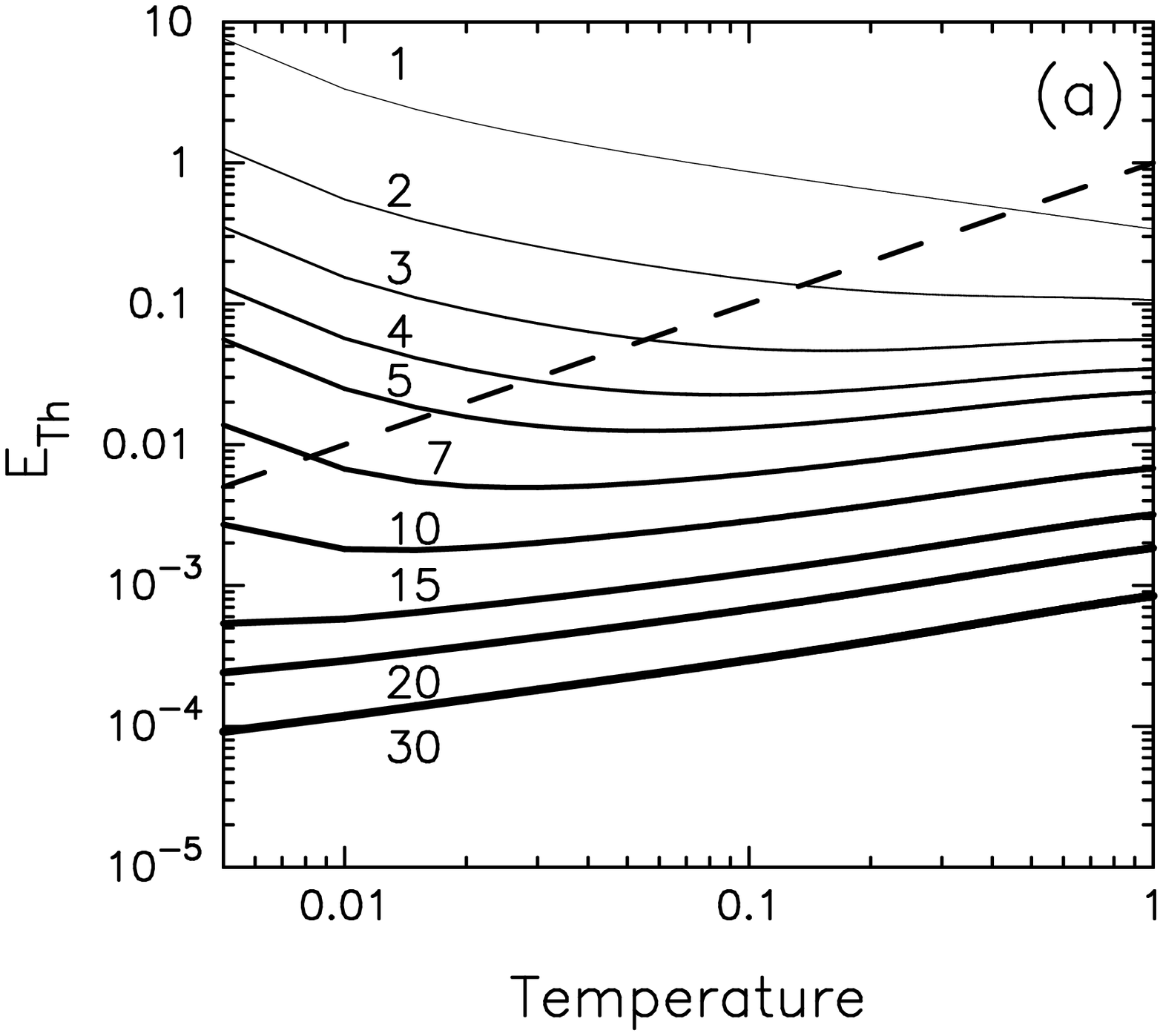}
\includegraphics[width=.465\textwidth]{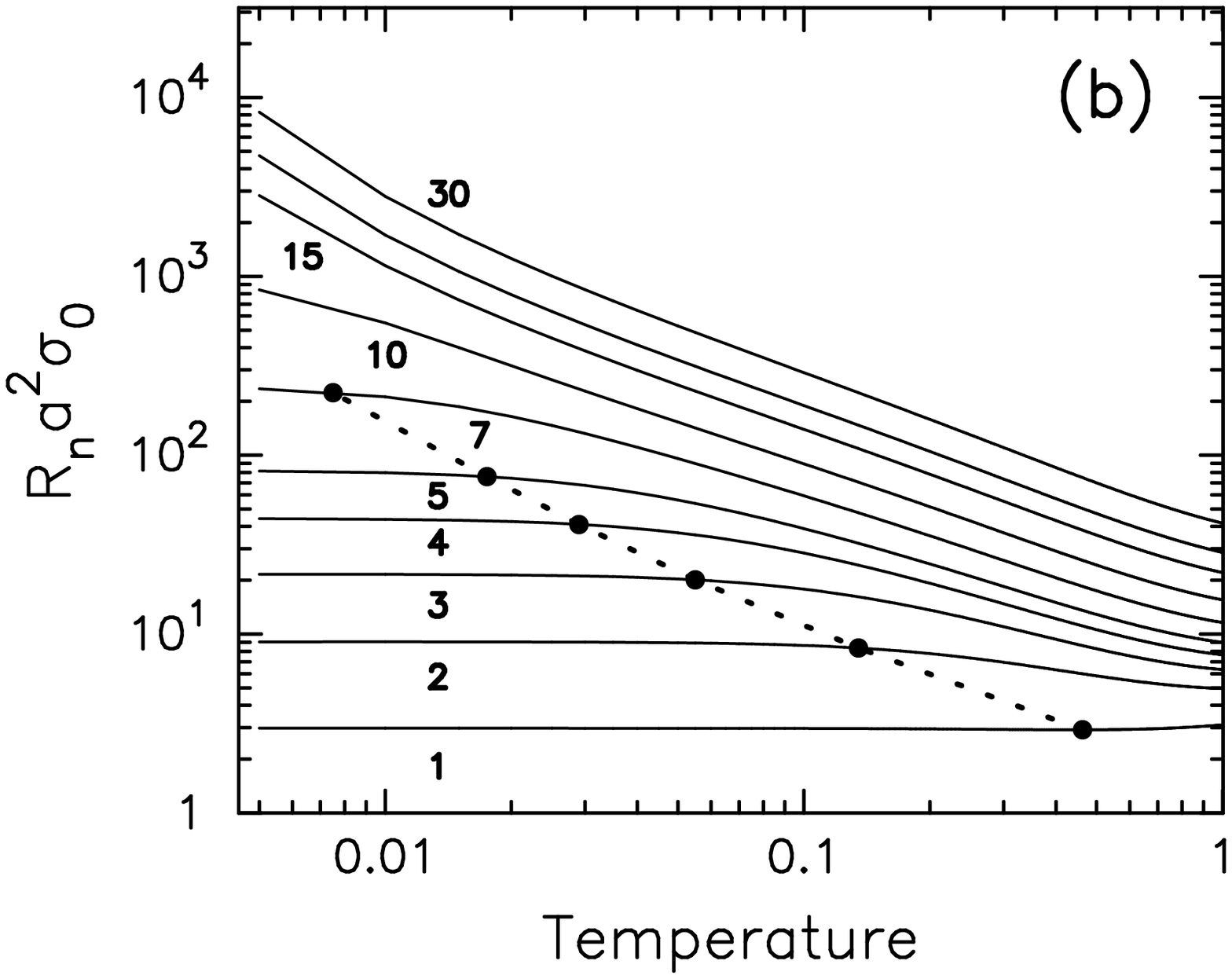}
}
\caption{(a) Thouless energy for the $U=5$ nanostructure as a function of
temperature on a log-log plot.  The different curves are for different 
thicknesses of the barrier.  The dashed line is the curve $E_{Th}=T$, and
the special points that denote the crossover from tunneling to incoherent
transport correspond to the points of intersection of the solid
lines with the dashed line. (b) Resistance-area product plotted versus the
temperature on a log-log plot.  The different curves correspond to different
barrier thicknesses.  The solid dots and the dotted line plots the
points where $E_{Th}=T$.  Note how the curves are flat at low temperature
and for thin junctions indicating tunneling (but the tunneling resistance
does not grow exponentially with the thickness when we are so close to the
metal-insulator transition and at finite temperature--note the unequal
spacing of the lines with $N$). At higher temperature,
the resistance picks up strong temperature dependence, and the transport
is best described by incoherent thermally activated processes. The numerical
labels on the figures denote the thickness of the barrier of the nanostructure
and the constant satisfies $\sigma_0=2e^2/ha^2$.
} \label{fig:3}
\end{vchfigure}

The generalized formula for the Thouless energy 
is~\cite{freericks_2004,freericks_unpub}
\begin{equation}
E_{Th}=\frac{\hbar}{R_na^22e^2\int d\omega [-df/d\omega] \rho_{bulk}(\omega) L},
\label{eq: thouless}
\end{equation}
where $a$ is the lattice spacing, $e$ is the electrical charge, $\hbar$ is
Planck's constant, $R_n$ is the junction resistance, $f(\omega)=
1/[1+\exp(\omega/T)]$ is the Fermi-Dirac distribution, 
$\rho_{bulk}$ is the interacting bulk DOS of the barrier material,
and $L=Na$ is the thickness of the barrier.

The resistance-area product
versus temperature is plotted in Fig.~\ref{fig:3}~(b).
Note the flat regions for low $T$ and thin junctions.  This is a signal that the
transport is dominated by tunneling, but because the gap is so small in
this system, we do not see an exponential increase in the resistance-area
product with the thickness of the junction.  Instead, it increases with a 
functional behavior in between that of an exponential increase and of a 
linear increase.  We also include a plot of the points where the Thouless
energy is equal to the temperature.  In this case, they are close to, but
not exactly at the point where the tunneling behavior crosses over to
incoherent transport (which has a strong temperature dependence).  As $U$
is increased further, this separation becomes more readily 
apparent~\cite{freericks_unpub}.
We believe this anomalous behavior occurs because the system is so close
to the critical point of the metal-insulator transition,

\section{Conclusions}

In this contribution we examined properties of a nanostructure composed
of metallic leads and a barrier that could be tuned through the 
metal-insulator transition (described by the Falicov-Kimball model).
We concentrated on general properties of the thin barrier (where
we showed it always has a metallic DOS generated by the normal proximity
effect with the metallic leads, although the weight within this metallic 
``subband'' can be quite small).  We also investigated the DOS and transport
properties of a near critical Mott insulating barrier with $U=5$ (the 
critical value of the transition on a simple cubic lattice is $U_c\approx 4.9$
within DMFT).  We found the system shows behavior that looks like tunneling,
but it also has a number of anomalies, the most important being that
the resistance does not grow exponentially with the junction thickness in
the tunneling regime.  We also defined a generalization of the Thouless energy 
that reduces to the ballistic and diffusive limits, but can also describe 
strongly correlated insulators.  We found the Thouless energy picks up
strong temperature dependence in this regime and the point where
$E_{Th}=T$ determines an approximate crossover from tunneling to
incoherent transport, where the resistance is proportional to the bulk
resistivity of the barrier multiplied by some geometrical factors (and
the resistivity has strong exponentially activated behavior in $T$ for
a correlated insulator).

There are a number of future directions that are important to consider as
one tries to examine models closer to experimental systems.  First, there
will be a charge transfer (electronic charge 
reconstruction)~\cite{nikolic_freericks_miller_2002,okamoto_millis_2004}
when the chemical potentials do not match, which can have strong effects
on correlated systems, especially correlated insulators, second, it
is useful to consider capacitive effects for these devices, since the
junction capacitance will play a role in the switching speed, and third,
it would be interesting to extend this analysis from equilibrium/linear response
to nonequilibrium/nonlinear response, where one could directly calculate
the current-voltage characteristic of the device, and determine the origins of
its nonlinearities.

\begin{acknowledgement}
We acknowledge support from the National
Science Foundation under grant number DMR-0210717 and the Office of Naval
Research under grant number N00014-99-1-0328. Supercomputer time was
provided by the Arctic Region Supercomputer Center and by the
Mississippi Region Supercomputer Center ERDC.
\end{acknowledgement}

\end{document}